\documentclass[fleqn,twoside]{article}
\usepackage{espcrc2}
\usepackage{graphicx}
\usepackage[figuresright]{rotating}
\newcommand{\MR}{m_{\cal R}}

\newcommand{\psip}{\psi(2S)}
\newcommand{\psipp}{\psi(3770)}
\newcommand{\jpsi}{J/\psi}
\newcommand{\DDbar}{D\overline{D}}
\newcommand{\EE}{e^+e^-}
\newcommand{\MM}{\mu^+\mu^-}

\newcommand{\pp}{\pi^+\pi^-}

\newcommand{\rhopi}{\rho\pi}
\newcommand{\omegapi}{\omega\pi^0}
\newcommand{\beq}{\begin{equation}}
\newcommand{\eeq}{\end{equation}}
\newcommand{\bfg}{\begin{figure}}
\newcommand{\efg}{\end{figure}}
\newcommand{\bitm}{\begin{itemize}}
\newcommand{\eitm}{\end{itemize}}
\newcommand{\bnum}{\begin{enumerate}}
\newcommand{\enum}{\end{enumerate}}
\newcommand{\btbl}{\begin{table}}
\newcommand{\etbl}{\end{table}}
\newcommand{\btbu}{\begin{tabular}}
\newcommand{\etbu}{\end{tabular}}
\newcommand{\aggg}{a_{3g}}
\newcommand{\agcc}{a_{\gamma}}
\newcommand{\agee}{a_c}
\newcommand{\rdr}{\sigma}
\newcommand{\rpdr}{\sigma^{\prime}}

\title{The Interference between Virtual Photon and $1^{--}$ Charmonium
in $\EE$ Experiment}
\author{P.~Wang  \address[IHEP]{Institute of High Energy Physics,
P.O.Box 918, Beijing 100039, China}
\thanks{Supported by 100 Talents Program of CAS (U-25)},
C.~Z.~Yuan \addressmark[IHEP],
X.~H.~Mo \addressmark[IHEP]$^,$\address[CCAST]{China Center of Advanced
Science and Technology, Beijing 100080, China} ,
D.~H.~Zhang \addressmark[IHEP]
}

\date{\today}
\begin{document}

\begin{abstract}

$\EE$ experiments producing charmonium are reviewed. It is
found that the contribution of the continuum amplitude via virtual
photon was neglected in almost all the experiments and the channels
analyzed. It is shown that the contribution of the continuum part may
affect the final results significantly in $\psip$ and $\psipp$ decays,
while the interference between continuum and resonance amplitudes
may even affect the $\jpsi$ decays as well as the $\psip$ and $\psipp$.
This should be considered in analyzing the ``$\rho\pi$ puzzle''
between $\jpsi$ and $\psip$ decays, and the difference between
inclusive hadron and $\DDbar$ cross sections in $\psipp$ decays.

\end{abstract}

\maketitle

\section{Introduction}\label{intro}

There are three well-known problems in the study of the
charmonium decays, namely the relative phase between
strong and electromagnetic amplitudes of the $1^{--}$
charmonium decays, ``$\rhopi$ puzzle'' between $\jpsi$
and $\psip$ decays, and non-$\DDbar$ decays of $\psipp$.

The attempt to understand the strong decays of $\jpsi$
via three-gluon and the electromagnetic decays via one-photon
annihilation reveals the relative phase between these two
amplitudes is close to $90^{\circ}$~\cite{suzuki,phyreport,phase,chen},
while for the radially excited $\psip$, the phase is
$0^\circ$ or $180^{\circ}$~\cite{suzuki,chen}. This indicates there
would be no interference between these two amplitudes in
$\jpsi$ decays, but strong interference in $\psip$ decays.

It was found that in $\psip$ hadronic decays,
some decay modes are abnormally suppressed compared with the
corresponding $\jpsi$ decays based on perturbative QCD (pQCD)
prediction. This suppression was first observed by the Mark-II
in vector pseudoscalar (VP) decay modes like $\rho\pi$ and
$K^*\overline{K}$~\cite{mk2}, and confirmed by BES~\cite{besVP}.
Moreover, BES also observed the suppression in
vector tensor (VT) decays of $\psip$~\cite{besVT}.
This has led to active theoretical efforts in solving the
problem~\cite{suzuki,chen,rhopitheo,rosner}. Unfortunately, most
of the models were ruled out by the experiments, while some
others need further experimental test.

There is a renewed interest in $\psipp$ studies because of the
upcoming high precision measurements by CLEO-c~\cite{cleoc}
and BES-III~\cite{bes3}. One of the
puzzling problems in $\psipp$ decays is that the $\DDbar$ cross
section may be significantly lower than the inclusive hadronic
cross section~\cite{nonddbar}. This is in contradiction with
the commonly accepted picture that $\psipp$ decays predominantly
to the OZI allowed $\DDbar$ states.

These three topics play important roles in understanding
the charmonium decay dynamics. In this paper we examine
what the experiments observe and what theories
analyze on charmonium produced in $\EE$ experiments.
We present a self-consistent analysis by
considering the unavoidable background process in
$\EE$ experiment, namely, the continuum process.
We show that, for exclusive decays of these charmonium
states, the contribution of this process could be very
important, or even if the direct contribution is relatively
small, the interference between this term and other dominant
amplitudes may contribute a non-negligible part.

\section{Experimentally observed cross section}

\begin{figure}
\begin{minipage}{8cm}
\includegraphics[width=3.25cm,height=2.5cm]{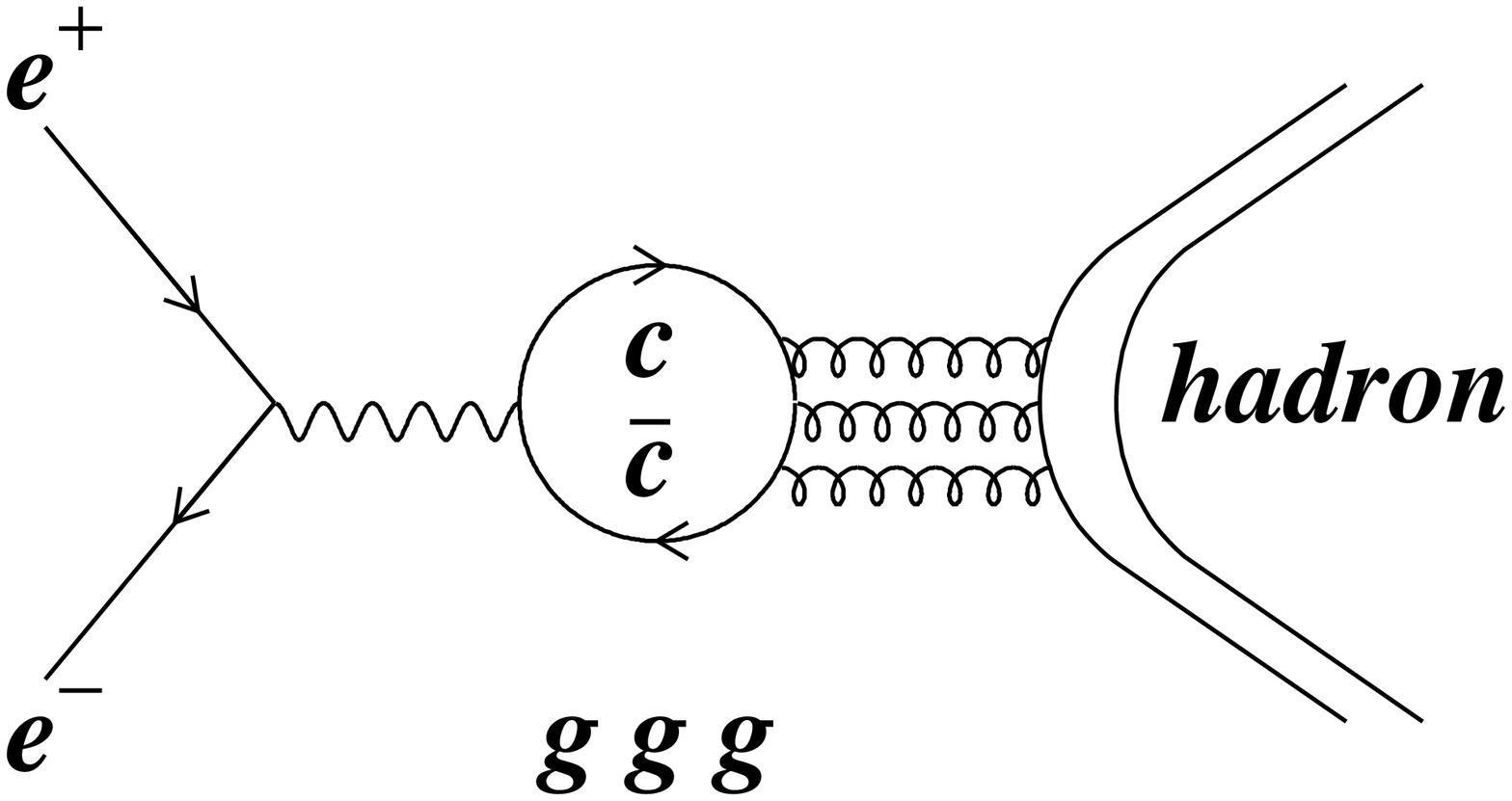}
\hskip 0.5cm
\includegraphics[width=3.25cm,height=2.5cm]{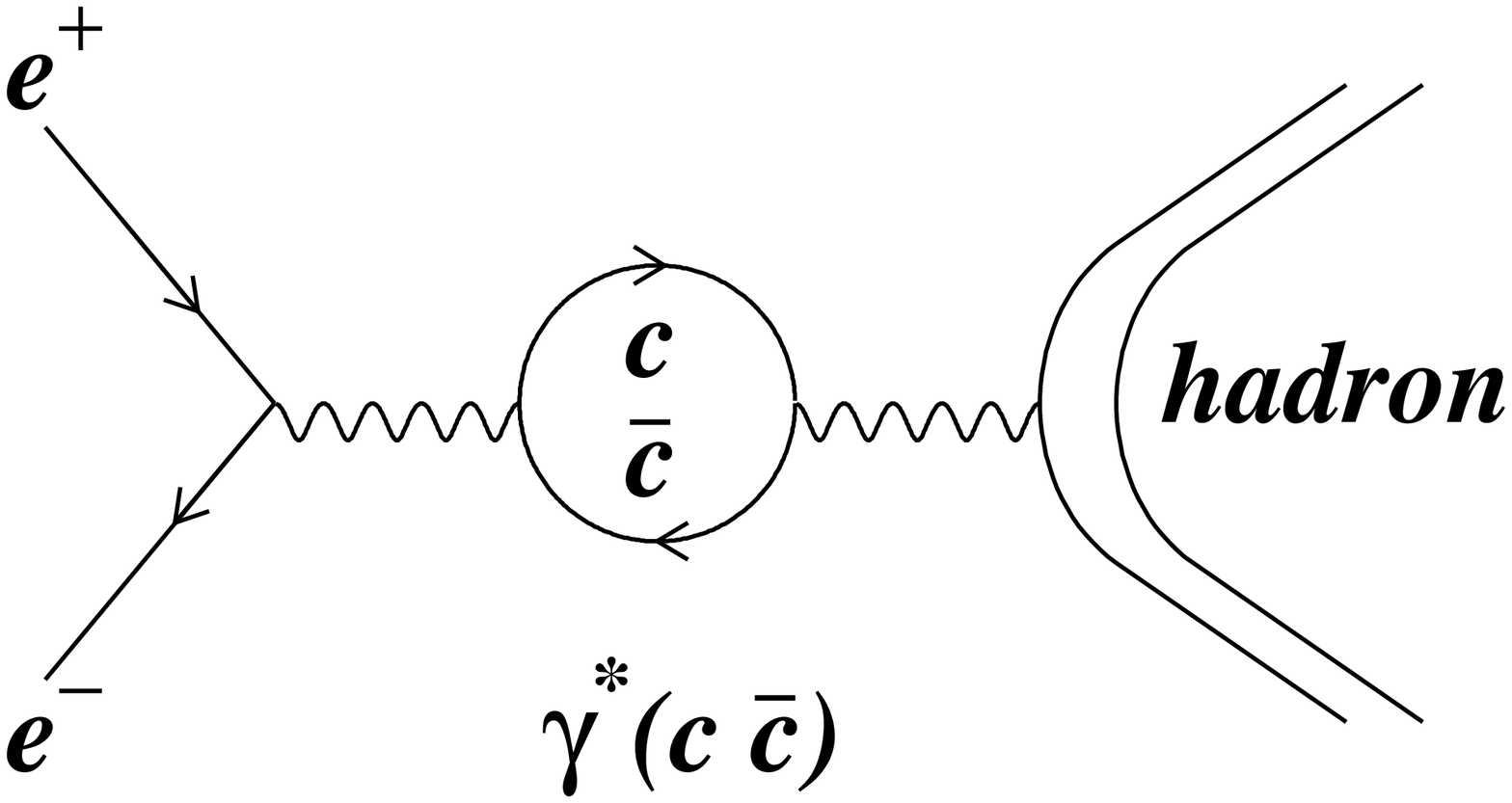}
\end{minipage}
\begin{minipage}{8cm}
\center
\includegraphics[width=3.5cm,height=2.5cm]{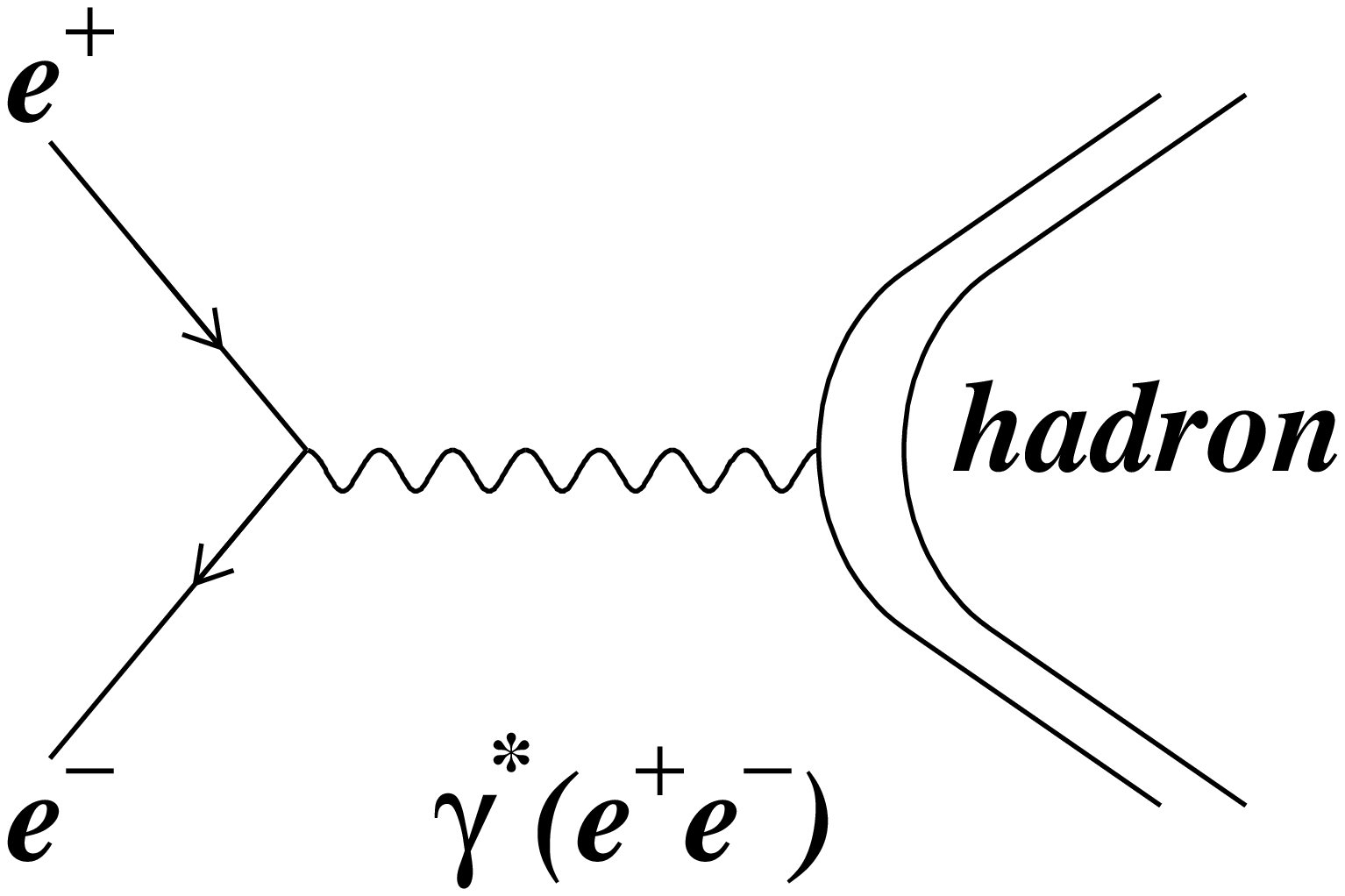}
\end{minipage}
\caption{\label{threefymn} The three classes of diagrams of
$\EE\rightarrow light\, \, hadrons$ at charmonium resonance. The
charmonium state is represented by a charm quark loop.}
\end{figure}

It is known that $\jpsi$ or $\psip$ decays into light hadrons via
strong and electromagnetic interactions. At the leading order in
$\alpha_s(m_c)$ and $1/m_c$,
it goes through three-gluon and one-photon annihilation of which
the amplitudes are denoted by $\aggg$ and $\agcc$
respectively~\cite{phyreport,haber}. This is also true for $\psipp$ in its
OZI suppressed decay into light hadrons.
In general, for the resonance ${\cal R}$ (${\cal R}=\jpsi$,
$\psip$ or $\psipp$),
the cross section at the Born order is expressed as
\beq
\rdr_{B}(s)= \frac{4\pi s\alpha^2}{3}|\aggg+\agcc|^2~,
\label{born}
\eeq
where $\sqrt{s}$ is the C.M. energy,
$\alpha$ is the fine structure constant.
If the $\jpsi$, $\psip$ or $\psipp$ is produced in $\EE$ collision,
the process
\begin{equation}
e^+e^- \rightarrow \gamma^* \rightarrow hadrons
\nonumber
\end{equation}
could produce the same final hadronic states as charmonium decays
do~\cite{rudaz}. We denote its amplitude by $\agee$, then the
cross section becomes \beq \rpdr_{B}(s) =\frac{4\pi s
\alpha^2}{3}|\aggg+\agcc+\agee|^2~. \label{bornp} \eeq So what
truly contribute to the experimentally measured cross section are
three classes of diagrams, $i.e.$ the three-gluon decays, the
one-photon decays, and the one-photon continuum process, as
illustrated in Fig.~\ref{threefymn}, where the charm loops stand
for the charmonium state, and the photons and gluons are highly
off-shell and can be treated perturbatively. To analyze the
experimental results, we must take into account three amplitudes
and two relative phases.

For an exclusive mode, $\agee$ can be expressed by
\begin{equation}
\agee(s) = \frac{{\cal F}(s)}{s} e^{i \phi^{\prime}}~,
\label{agee}
\end{equation}
where
$\phi^\prime$ is the phase relative to $\aggg$;
${\cal F}(s)$ depends on the individual mode, and
for simplicity, the phase space factor is incorporated into
$|{\cal F}(s)|^2$.
The one-photon annihilation amplitude can be written as
\begin{equation}
\agcc(s)=\frac{3\Gamma_{ee}{\cal F}(s)/(\alpha \sqrt{s})}
{s-\MR^2+i \MR \Gamma_t}e^{i\phi}~,
\label{agcc}
\end{equation}
where $\MR$ and $\Gamma_t$ are the mass and the total width of
${\cal R}$, $\Gamma_{ee}$ is the partial width to $\EE$,
$\phi$ is the phase relative to $\aggg$.
The strong decay amplitude $\aggg$ is defined by
${\cal C}\equiv |\aggg/\agcc|$, which is the relative
strength to $\agcc$, so
\begin{equation}
\aggg(s) = {\cal C} \cdot \frac{3\Gamma_{ee}{\cal F}(s)/(\alpha \sqrt{s})}
{s-\MR^2+i \MR \Gamma_t}~.
\label{aggg}
\end{equation}
For resonances, ${\cal C}$ can be taken as a constant.

\begin{table*}[htb]
\caption{\label{estimation} Estimated amplitudes at $\jpsi$, $\psip$
and $\psipp$ peaks.}
\vskip 0.2 cm
\center
\btbu{r||ccc} \hline \hline
$\sqrt{s}$~~~~~~~~  & $m_{\jpsi}$   &   $m_{\psip}$    & $m_{\psipp}$
\\ \hline \hline
$| \aggg(m^2_{\cal R}) |^2 \propto
$& 70\% $\sigma^{\jpsi}_B$  & 19\% $\sigma^{\psip}_B$
      & $\sim$ 1\% $\sigma^{\psipp}_B$  \\
$| \agcc(m^2_{\cal R}) |^2 \propto
$& 13\% $\sigma^{\jpsi}_B$  & 1.6\% $\sigma^{\psip}_B$
& $2.5\times 10^{-5} \sigma^{\psipp}_B$   \\
$| \agee(m^2_{\cal R}) |^2 \propto $& 20~nb & 14~nb & 14~nb  \\ \hline \hline
\etbu
\end{table*}

In principle, $\aggg$, $\agcc$ and $\agee$
depend on individual exclusive mode both in absolute values and in
relative strengths. In this paper, for illustrative purpose,
following assumptions are used for an exclusive hadronic mode:
${\cal F}(s)$ is replaced by $\sqrt{R(s)}$, where $R(s)$ is the
ratio of the inclusive hadronic cross section to the $\MM$
cross section measured at nearby energy \cite{besR};
in Eq.~(\ref{aggg}),
\beq
{\cal C}=
\sqrt{\frac{B({\cal R}\rightarrow ggg\rightarrow hadrons)}
{B({\cal R}\rightarrow \gamma^* \rightarrow hadrons) }}~.
\label{cratio}
\eeq
Here $B({\cal R}\rightarrow \gamma^* \rightarrow hadrons)
=B_{\MM}R(s)$, where $B_{\MM}$ is the $\MM$ branching ratio; while
$B({\cal R}\rightarrow ggg \rightarrow hadrons)$ is calculated as
following:  we first estimate the branching ratio of
$B({\cal R}\rightarrow \gamma gg)+B({\cal R}\rightarrow ggg)$
by subtracting the lepton pairs, $\gamma^* \rightarrow hadrons$,
and the modes with charmonium production from the total branching ratio
(100\%). Then using pQCD result~\cite{guali}
$B({\cal R}\rightarrow \gamma gg)/B({\cal R}\rightarrow ggg)\approx 6 \%$
we obtain $B({\cal R}\rightarrow ggg \rightarrow hadrons)$.
Table~\ref{estimation} lists all the estimations
used as inputs in the calculations, where $\sigma^{\cal R}_B$ is
the total resonance cross section of Born order at $s=\MR^2$
obtained from
\beq
\sigma_0^{\cal R}(s)=\frac{12\pi\Gamma_{ee}\Gamma_t}
{(s-\MR^2)^2+\MR^2\Gamma_t^2}~.
\label{sgmzeror}
\eeq

The cross section by $\EE$ collision incorporating radiative
correction on the Born order is expressed by~\cite{rad}
\begin{equation}
\sigma_{r.c.} (s)=\int \limits_{0}^{x_m} dx
F(x,s) \frac{\sigma_{0}(s(1-x))}{|1-\Pi (s(1-x))|^2}~,
\label{radsec}
\end{equation}
where $\sigma_{0}$ is $\rdr_{B}$ or $\rpdr_{B}$ by Eq.~(\ref{born})
or (\ref{bornp}),
$F(x,s)$ has been calculated in Ref.~\cite{rad}
and $\Pi (s)$ is the vacuum polarization factor~\cite{vacuum};
the upper limit of the integration $x_m=1-s_m/s$
where $\sqrt{s_m}$ is the experimentally required
minimum invariant mass of the final state $f$ after
losing energy to multi-photon emission. In this paper, we
assume that $\sqrt{s_m}$ equals to $90\%$ of the resonance
mass, $i.e.$ $x_m=0.2$.

For narrow resonances like $\jpsi$ and $\psip$, one should consider
the energy spread function of $e^+e^-$ colliders:
\begin{equation}
G(\sqrt{s},\sqrt{s'})=\frac{1}{\sqrt{2 \pi} \Delta}
          e^{ -\frac{(\sqrt{s}-\sqrt{s'})^2}{2 {\Delta}^2} },
\label{spread}
\end{equation}
where $\Delta$ describes the C.M. energy spread of the accelerator,
$\sqrt{s}$ and $\sqrt{s'}$ are the nominal and actual C. M.
energy respectively. Then the experimentally measured cross section
\begin{equation}
\sigma_{exp} (s)=\int \limits_{0}^{\infty}
   \sigma_{r.c.} (s') G(\sqrt{s},\sqrt{s'}) d\sqrt{s'}~.
\label{expsec}
\end{equation}

The radiative correction reduces the maximum cross sections of $\jpsi$,
$\psip$ and $\psipp$ by $52\%$, $49\%$ and $29\%$ respectively.
The energy spread further reduces the cross sections of
$\jpsi$ and $\psip$ by an order of magnitude. The radiative correction
and energy spread also shift the maximum height of the resonance peak
to above the resonance mass. Take $\psip$ as an example,
from Eq.~(\ref{sgmzeror}), $\sigma_B^{\psip}=7887$~nb at $\psip$
mass; substitute $\sigma_0(s)$ in Eq.~(\ref{radsec}) by
$\sigma_0^{\cal R}(s)$ in Eq.~(\ref{sgmzeror}), $\sigma_{r.c.}$
reaches the maximum of $4046$~nb at $\sqrt{s}=m_{\psip}+9$~keV;
with the energy spread $\Delta=1.3$~MeV at BES/BEPC, combining
Eqs.~(\ref{sgmzeror}$-$\ref{expsec}), $\sigma_{exp}$ reaches
the maximum of $640$~nb at $\sqrt{s}=m_{\psip}+0.14$~MeV.
Similarly, at $\jpsi$, with BES/BEPC energy spread $\Delta=1.0$~MeV,
the maximum of $\sigma_{exp}$ is $2988$~nb. At DORIS,
the maximum of $\sigma_{exp}$ at $\jpsi$ is $2190$~nb
($\Delta$= 1.4~MeV), and at $\psip$, it is $442$~nb
($\Delta$= 2.0~MeV). In this paper, we calculate $\sigma_{exp}$
at the energies which yield the maximum inclusive hadronic
cross sections.

To measure an exclusive mode in $\EE$ experiment,
the contribution of the continuum part should be
subtracted from the experimentally measured $\sigma^{\prime}_{exp}$
to get the physical quantity $\sigma_{exp}$, where
$\sigma_{exp}$ and $\sigma^{\prime}_{exp}$ indicate the experimental
cross sections calculated from
Eqs.(\ref{radsec}$-$\ref{expsec})
with the substitution
of $\rdr_{B}$ and $\rpdr_{B}$ from Eqs.~(\ref{born}) and (\ref{bornp})
respectively for $\sigma_{0}$ in Eq.~(\ref{radsec}).
Up to now, most of the measurements did not include this contribution
and $\sigma^{\prime}_{exp}=\sigma_{exp}$ is assumed at least at
$J/\psi$ and $\psip$.
As a consequence, the theoretical analyses are based on
$\sigma_{exp}$, while the experiments
actually measure $\sigma^{\prime}_{exp}$.

We display the effect from the continuum amplitude and corresponding
phase for $\jpsi$, $\psip$ and $\psipp$ respectively. To do this,
we calculate the ratio
\beq
k(s) \equiv \frac{ \sigma^{\prime}_{exp}(s) - \sigma_{exp}(s) }
              { \sigma^{\prime}_{exp}(s) }
\label{ratiok}
\eeq
\begin{figure}
\includegraphics[width=7.5cm,height=4cm]{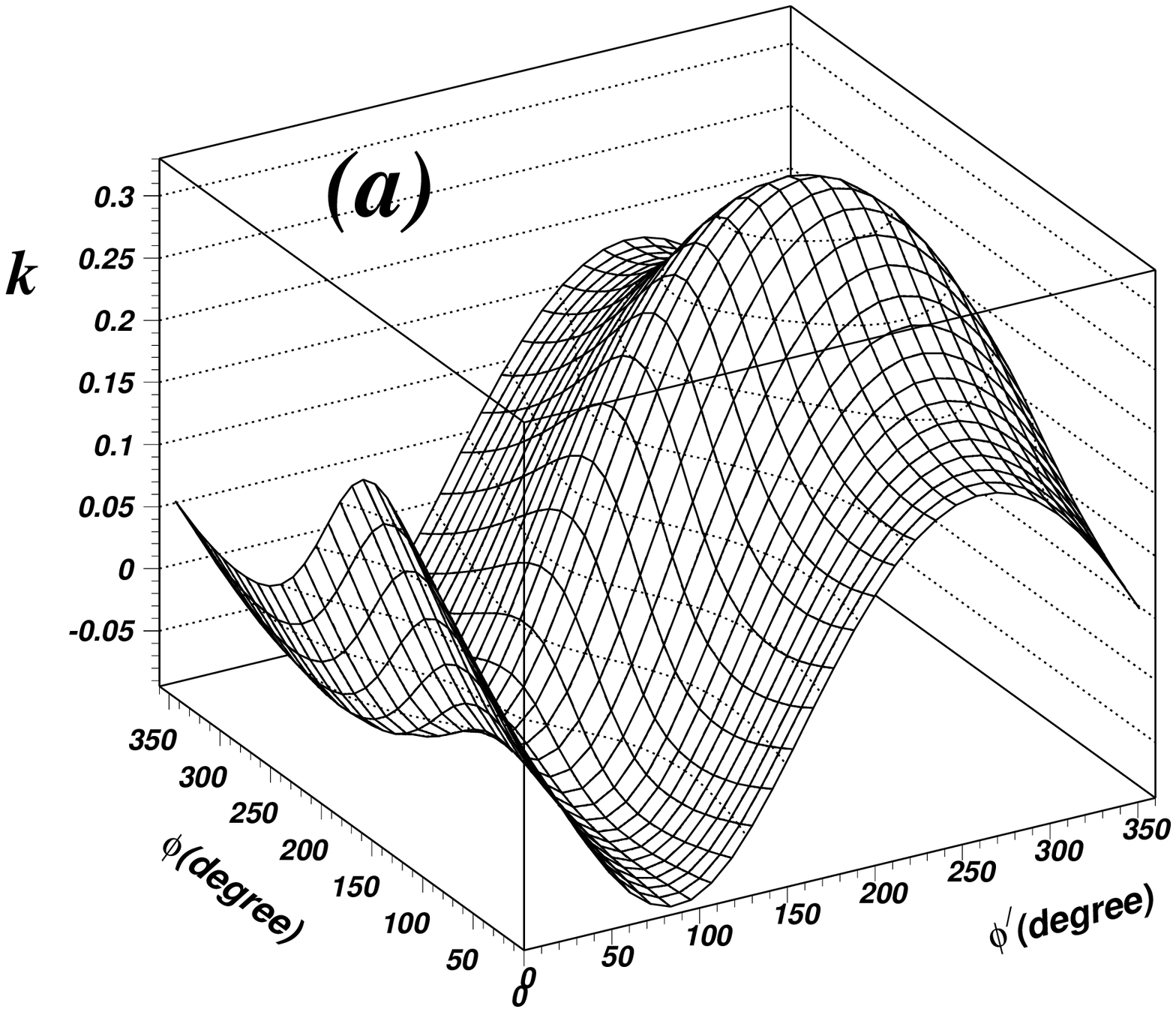}
\vskip 0.5cm
\includegraphics[width=7.5cm,height=4cm]{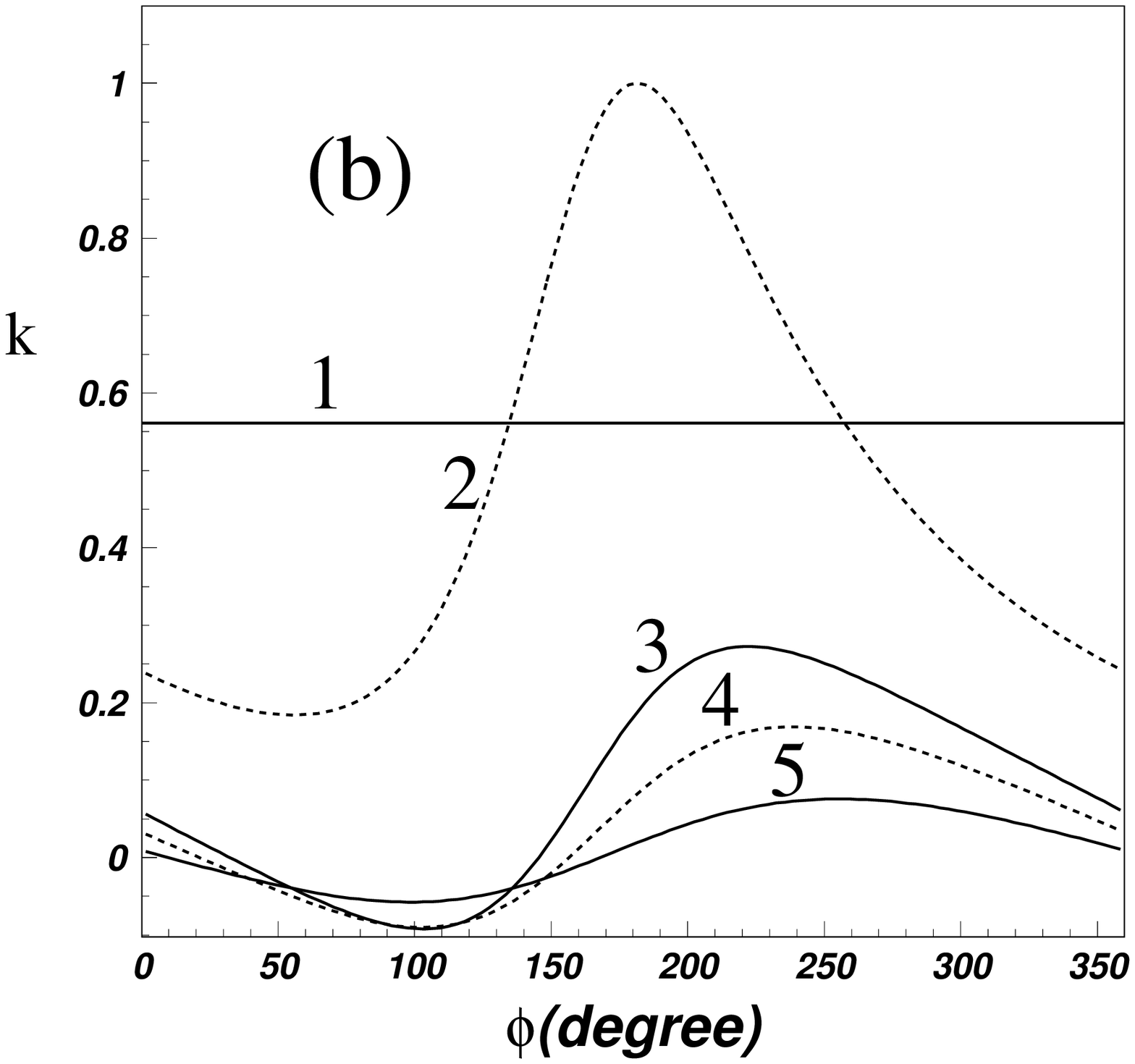}
\caption{\label{kratio} (a) $k$ as a function of
$\phi$ and $\phi^{\prime}$ for $\psip$,
with input from Table~\ref{estimation},
and (b) $k$ as a function of $\phi$ ($\phi = \phi^{\prime}$)
for different ratios of $|\aggg|$ to $|\agcc|$: line 1 to 5
for $\aggg=0$, $|\aggg|=|\agcc|$, $|\aggg|=3.4|\agcc|$,
$|\aggg|=5|\agcc|$ and $|\aggg|=10 |\agcc|$, respectively.}
\end{figure}
as a function of $\phi$ and $\phi^{\prime}$,
as shown in Fig.~\ref{kratio}a for $\psip$ at
$\sqrt{s}=m_{\psip}+0.14$~MeV for $\Delta=1.3$~MeV.
It can be seen that for certain values of the two phases,
$k$ deviates from 0, or equivalently the ratio
$\sigma^{\prime}_{exp}/ \sigma_{exp}$ deviates from 1,
which demonstrates that the continuum amplitude is non-negligible.
By assuming there is no extra phase between $\agcc$ and $\agee$ ($i.e.$
set $\phi=\phi^{\prime}$), we also work out the $k$ values for different
ratios of $|\aggg|$ to $|\agcc|$, as shown in Fig.~\ref{kratio}b:
line 3 corresponds to the numbers listed in
Table~\ref{estimation}, line 1 is for pure electromagnetic decay
channels, and others are chosen to cover the other
possibilities of the ratio $|\aggg|$ to $|\agcc|$.

\section{Continuum contribution for charmonium decay}\label{sct:cccd}
We now discuss separately the effect of continuum amplitude
for $\psipp$, $\psip$ and $\jpsi$.

At $\psipp$, the maximum resonance cross section of inclusive hadrons is
8~nb which predominantly decays into $\DDbar$, while the continuum cross
section is 14~nb which mainly goes to light hadrons.
Assuming 1\% of $\psipp$ decays to non-$\DDbar$ interferes
with the continuum amplitude, it could bring an effect of
maximum 1.9~nb in the observed cross section.
Such large constructive interferences could be responsible for
the larger cross section of inclusive hadrons by direct measurement of
$\EE \rightarrow \psipp \rightarrow hadrons$ than the $\DDbar$
cross section~\cite{nonddbar}.
As to the exclusive decays, it could make some of the decay modes
with small branching ratios more observable at the resonance.
For example,
if ${\cal B}(\psipp\rightarrow \rhopi) \approx 4\times10^{-4}$
(or equivalently, $\sigma_{\psipp\rightarrow \rho\pi} \approx 0.003$~nb)
as suggested in Ref.~\cite{rosner}, and
$\sigma(\EE\rightarrow\rhopi)\approx 0.014$~nb
at Born order by the model of Ref.~\cite{Achasov},
then the maximum interference could be 0.011~nb,
much larger than the pure contribution from $\psipp$ decays.

For $\psip$, as can be seen in Fig.~\ref{kratio}, the ratio
$\sigma_{exp}^{\prime}/\sigma_{exp}$ could deviate from $1$
substantially.
In general, $\aggg$, $\agcc$ and $\agee$ are different for different
exclusive mode, so $k$ could be different. This must be taken into
account in the fitting of $\agcc$, $\aggg$ and the phase in between.
It is noticeable that the observed
cross sections of some electromagnetic processes, such as
$\psip \rightarrow \pp$, $\omegapi$,
and the famous puzzling $\psip\rightarrow \rhopi$,
are three to four orders of magnitude smaller than
the total hadronic cross section of the continuum process, which is
about 14~nb.  Form factor estimation~\cite{fofa1}
gives these cross sections at continuum comparable to the ones
measured at the resonance~\cite{wym}. It implies that a substantial
part of the experimentally measured cross section could come from
the continuum amplitude $\agee$ instead of the $\psip$ decays, and
interference between these two amplitudes may even affect the
measured quantity further.  Therefore it is essential to measure the
production rate of $\pp$, $\omegapi$ and $\rhopi$ at the
continuum in order to get the correct branching ratios
of the $\psip$ decays. The same holds for VT decays
of $\psip$.

As for $\jpsi$, the interference between the amplitude $\agee$
and the resonance is at the order of a few percent.
It is smaller than the statistical
and systematic uncertainties of current measurements.
Nevertheless, for future high precision experiments such as
CLEO-c~\cite{cleoc} and BES-III~\cite{bes3}, when the
accuracy reaches a few per mille or even smaller level, it
should be taken into account.

\section{Dependence on experimental conditions}\label{sct:dexp}
Here we emphasize the dependence of the
observed cross section in $\EE$ collision on the experimental
conditions. The most crucial ones are the
accelerator energy spread and the beam energy setting
for the narrow resonances like $\jpsi$ and $\psip$.

\begin{figure}[hbt]
\begin{center}
\includegraphics[width=7.5cm,height=7.cm]{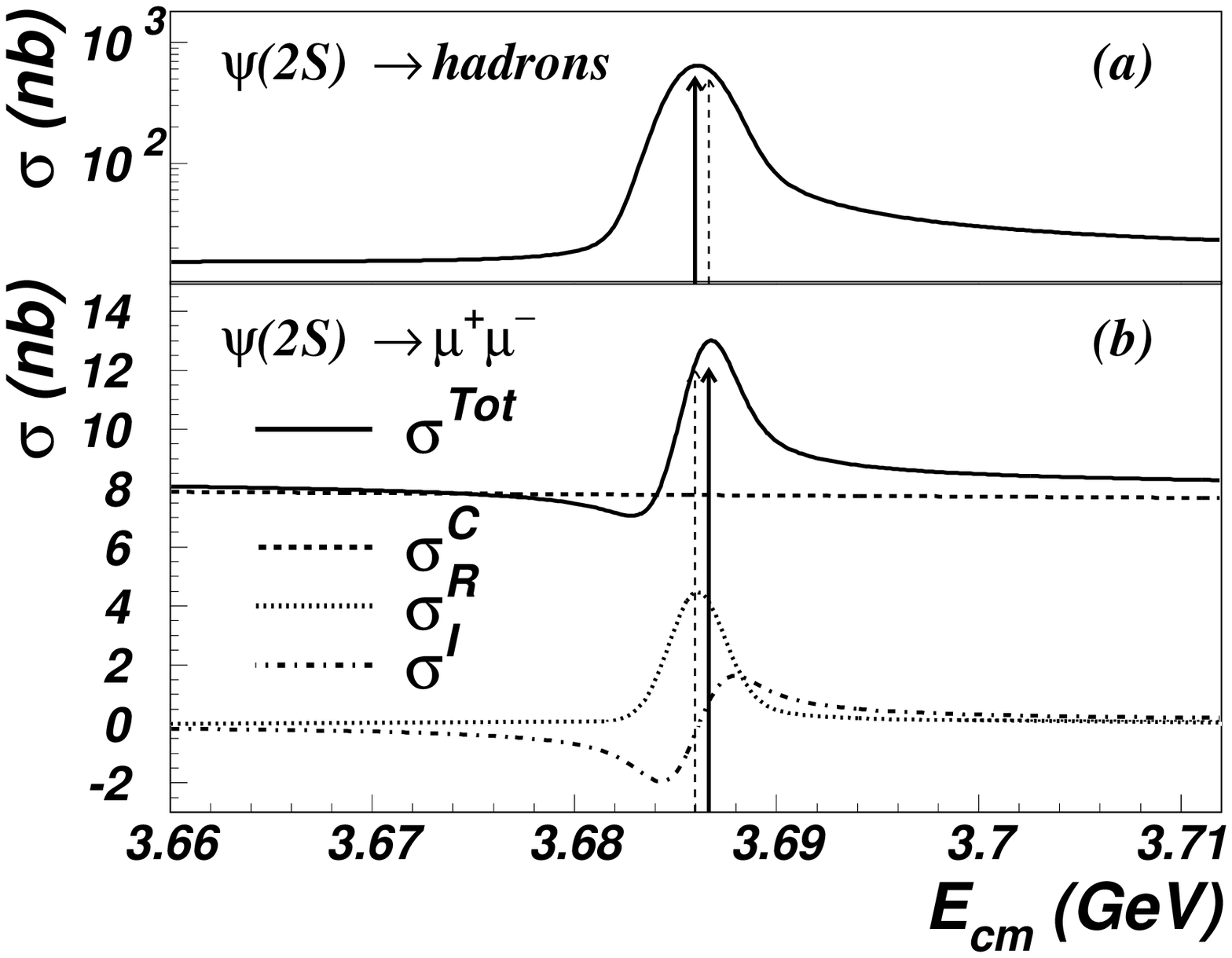}
\caption{\label{cmphup} Cross sections in the vicinity of $\psip$
for inclusive hadrons (a) and $\MM$ (b) final states.
The solid line with arrow indicates the peak position
and the dashed line with arrow the position of the other
peak. In (b), dashed line for QED continuum ($\sigma^C$),
dotted line for resonance ($\sigma^R$),
dash dotted line for interference($\sigma^I$), and
solid line for total cross section($\sigma^{Tot}$).}
\end{center}
\end{figure}

Fig.~\ref{cmphup} depicts the observed cross sections of
inclusive hadrons and $\mu^+\mu^-$ pairs at $\psip$ in actual
experiments. Two arrows in the figure denote the different positions of
the maximum heights of the cross sections. The height is reduced and the
position of the peak is shifted due to the radiative correction
and the energy spread of the collider. However, the energy smear hardly
affects the continuum part of the cross section.
The $\MM$ channel is further affected by the interference
between resonance and continuum amplitudes. As a consequence,
the relative contribution of the resonance and the continuum
varies as the energy changes.
In actual experiments, data are naturally taken at the energy
which yields the maximum inclusive hadronic cross section.
This energy does not coincide with the maximum cross section of each
exclusive mode. So it is important to know the beam spread and
beam energy precisely, which are needed in the delicate task to
subtract the contribution from $\agee$.

It is worth noting that in principle if $\agee$ is not considered
correctly, different experiments will give different results
for the same quantity, like the exclusive branching ratio
of the resonance, due to the dependence on beam energy spread
and beam energy setting.  The results will also be different for
different kinds of experiments, such as production of $\jpsi$
and $\psip$ in $p \bar{p}$ annihilation, or in $B$ meson decays.
This is especially important
since the beam spreads of different accelerators are much
different~\cite{phyreport} and charmonium results are expected
from $B$-factories.

\section{Summary and perspective}\label{sct:sum}
In summary, the continuum amplitude $\agee$, by itself or
through interference with the resonance, could contribute significantly
to the observed cross sections in $\EE$ experiments on charmonium physics.
Its treatment depends sensitively on the experimental details,
which has not been fully addressed in both
$\EE$ experiments and theoretical analyses.
So far, most of the measurements have large statistical
and systematic uncertainties, so this problem has been outside
the purview of concern. Now with large $\jpsi$
and $\psip$ samples from BES-II~\cite{bes2data}
and forthcoming high precision experiments CLEO-c~\cite{cleoc} and
BES-III~\cite{bes3}, the effect of $\agee$ needs to be treated
properly. To study it, the most promising way is to do energy scan
for every exclusive mode in the vicinity of the resonance, so that
both the amplitudes and the relative phases could be fit
simultaneously. In case this is not practicable, data sample
off the resonance with comparable integrated luminosity
as on the resonance should be collected to measure $|\agee|$,
which could give an estimation of its contribution to the
decay modes studied.
The theoretical analyses based on current available $\EE$ data,
particularly on $\psip$ may need to be revised correspondingly.


\end{document}